# Performance analysis and optimization of the JOREK code for many-core CPUs

T. B. Fehér[1], M. Hölzl[1], G. Latu[2], G.T.A. Huijsmans[2]

[1]Max Planck Institute for Plasma Physics, 85748 Garching, Germany
[2]CEA, IRFM, 13108, Saint-Paul-lez-Durance, France

This report investigates the performance of the JOREK code on the Intel Knights Landing and Skylake processor architectures. The OpenMP scaling of the matrix construction part of the code was analyzed and improved synchronization methods were implemented. A new switch was implemented to control the number of threads used for the linear equation solver independently from other parts of the code. The matrix construction subroutine was vectorized, and the data locality was also improved. These steps led to a factor of two speedup for the matrix construction.

# 1. Introduction

The JOREK code (Huysmans and Czarny 2007) is a nonlinear extended MHD code that can resolve toroidal X-point geometries. Its key application areas are the simulation of Edge Localized Modes (ELM) and disruptions. It is the most important numerical tool in Europe to study MHD instabilities in realistic tokamak geometry. JOREK has been granted a large amount of CPU hours on the Knights Landing (KNL) partition of the Marconi-Fusion supercomputer. The aim of the JOKLA project is to improve the performance of the JOREK code on the KNL architecture and in general for many-core CPUs.

The two main challenges are the efficient usage of the wide vector registers and the scaling over a large number of threads. We should note that these are not unique requirements for the KNL architecture. In fact, the Skylake nodes on Marconi have an identical vector register size, and the number of cores per node is also comparable. Because of these similarities, we will test and optimize the code for both the KNL and the Skylake architectures.

# 2. Profiling

The first phase of the project started with a detailed profiling on both the KNL and the Skylake partitions. We used an ASDEX Upgrade ELM simulation setup (Hölzl 2018). The main parameters for the test case are listed in Table 1. We have two variants: a small (*ntor*=3) and a medium (*ntor*=17) size test case. The benchmark was prepared by running the test for 2000 time steps. Using the restart files of this simulation we then performed three time steps in the nonlinear phase of the ELM crash, and used measurements from these three steps to analyze the code performance.

| Resolution | | JOREK parameters | | |
|---|---|---|---|---|
| | | | small | medium |
| n_tht | 173 | model | 303 | 303 |
| n_radial | 120 | N_tor | 3 | 17 |
| n_open | 10 | N_period | 8 | 1 |
| n_pol | 160 | N_plane | 4 | 32 |
| n_leg | 18 | MPI tasks | 2 | 18 |
| n_private | 8 | Compute nodes | 1 | 18 |

**Table 1** Main parameters of the AUG ELM test case

We compared the execution time of different code regions in the nonlinear phase of the simulation (after time step 2000). Fig. 1 shows the execution time of different parts of the computation on Skylake (red) and KNL (green) for the small test case. At the first step after the restart, we have an LU decomposition which dominates the execution time. In subsequent time steps, the LU decomposition is not repeated (unless the convergence becomes bad); instead the factorized matrix is used as a preconditioner. The right side of Fig. 1 shows the execution time of the third time step after the restart. At this time step the matrix construction and the iterative solver take the largest share of the execution time. During the first time step the SKL node is 2.9 times faster than the KNL node, later the ratio becomes 2.2.

Similar measurements were performed in the linear phase of the simulation (time steps 500–503). The results are very similar to the nonlinear case, only the execution time of the GMRES solver differs. In the following we will focus on the nonlinear phase of the simulation.

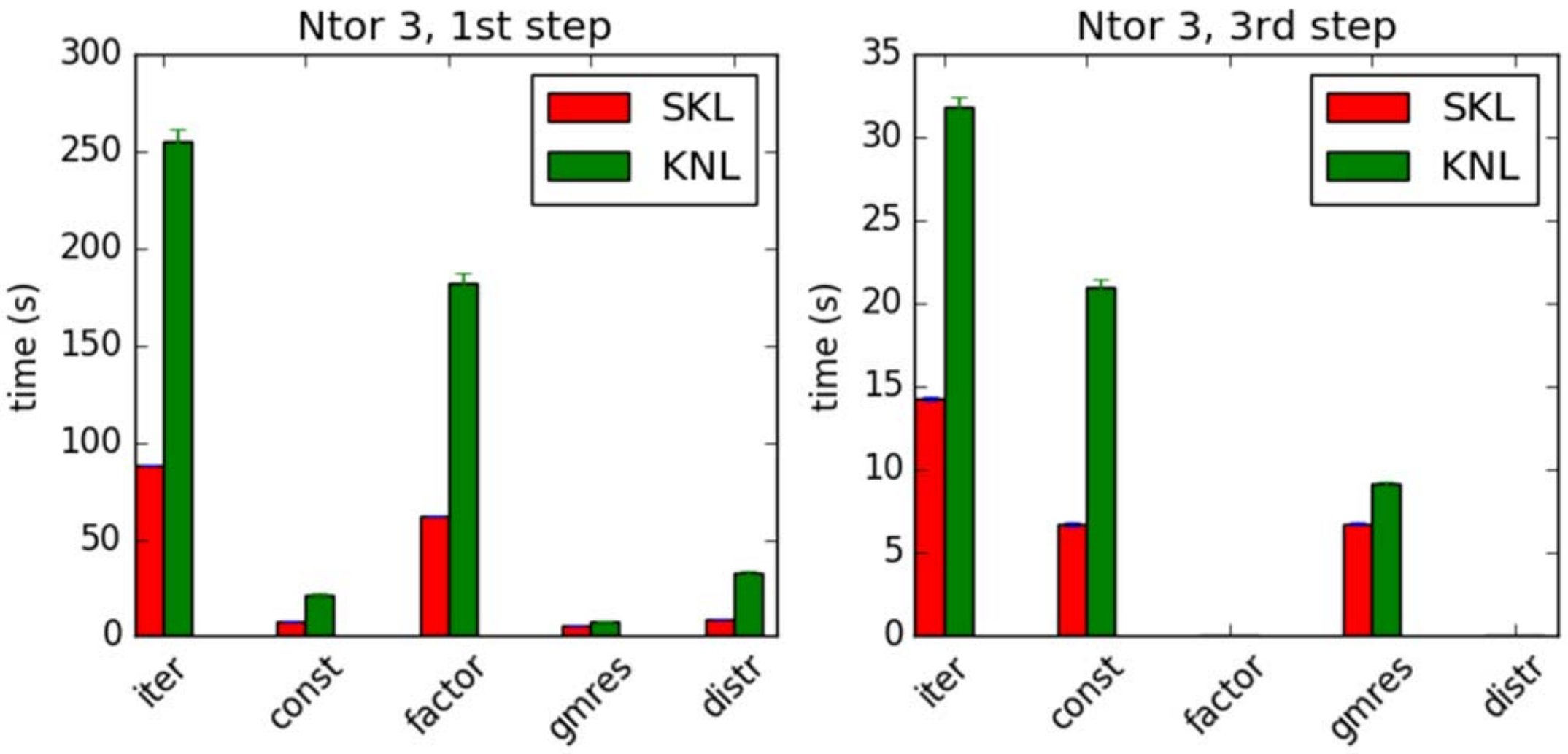

**Fig. 1** JOREK small test case execution time: first time step (left) and third time step (right) after restart. Red bars show the execution time on Skylake, while green bars show the execution time on Knights Landing. The first group of bars denote the total time of an iteration, which can be decomposed into matrix construction (const), LU factorization (factor), iterative solver (GMRES) and matrix distribution (distr).

# 3. **Compiler options**

The JOREK code uses the PaStiX (Parallel Sparse Matrix) library as a linear solver. The PaStiX library depends on the Scotch package to calculate the matrix ordering (permutation of the matrix to reduce fill-ins). We used the following optimization options during compilation of all three packages: *-O3 -no-prec-div -xCORE-AVX512 -mtune=Skylake.* For the compilation of the JOREK code, we additionally used the *-align array64byte* option.

We tested whether changing the optimization level or changing the precision of divisions would improve the execution time. Fig. 2 shows that the default settings (*-O3*) were already optimal.

Due to a compiler bug, initially it was not possible to test the effects of interprocedural optimizations among separate files (*-ipo* flag). This problem was reported to Intel, and it was later fixed in update 3 of the Intel 2018 compiler suite. With the latest compiler version, it was found that overall the *–ipo* flag does not improve the performance.

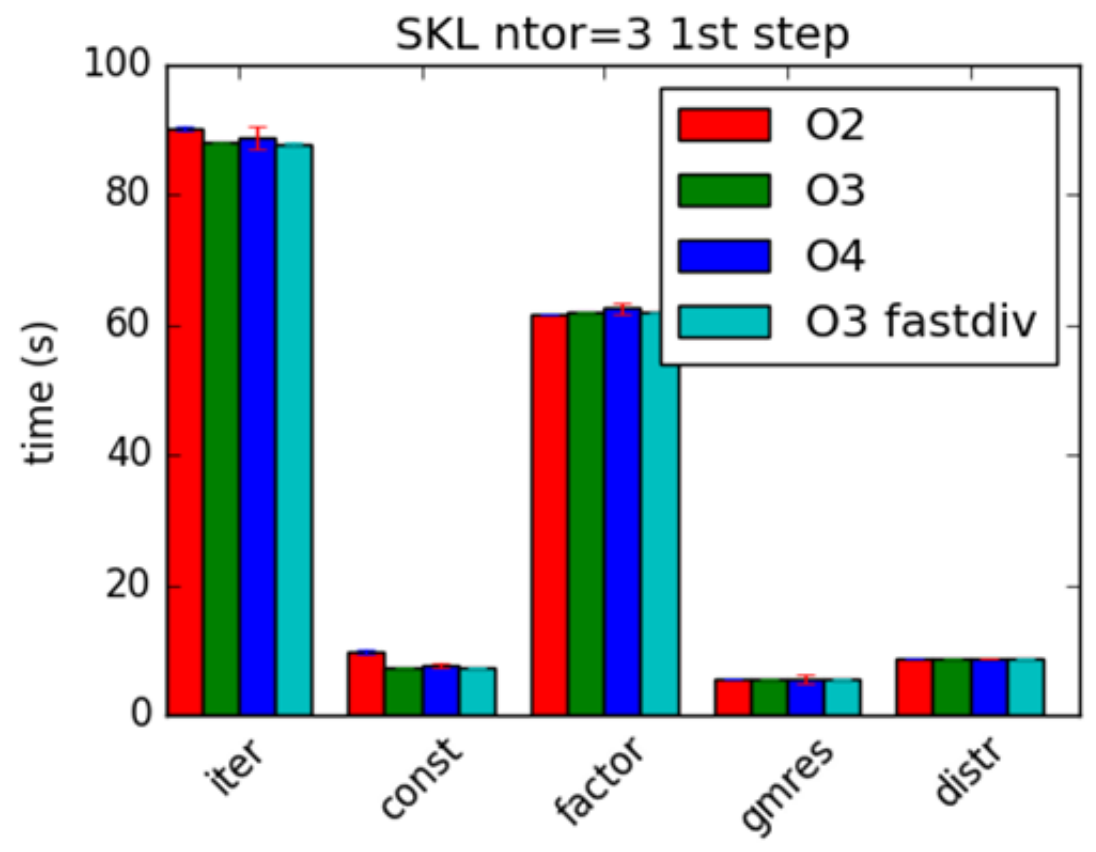

**Fig. 2** Execution times using different compiler flags

# 4. OpenMP scaling

## 4.1. *Skylake*

The OpenMP scaling of different parts of the code was tested. In this section we present strong scaling tests, where the amount of work is kept fixed as we increase the number of threads. First, we focus on the matrix construction. While the small test case scaled almost ideally (see Fig. 3), the large test case did not scale above four threads. Using the VTune Amplifier tool, it was identified that a critical section in the *construct_matrix* subroutine is responsible for the scaling problem. This subroutine has a large critical section around the loops which update the following global variables: `irn_glob(ilarge2)`, `jcn_glob(ilarge2)`, `A_glob(ilarge2)`.

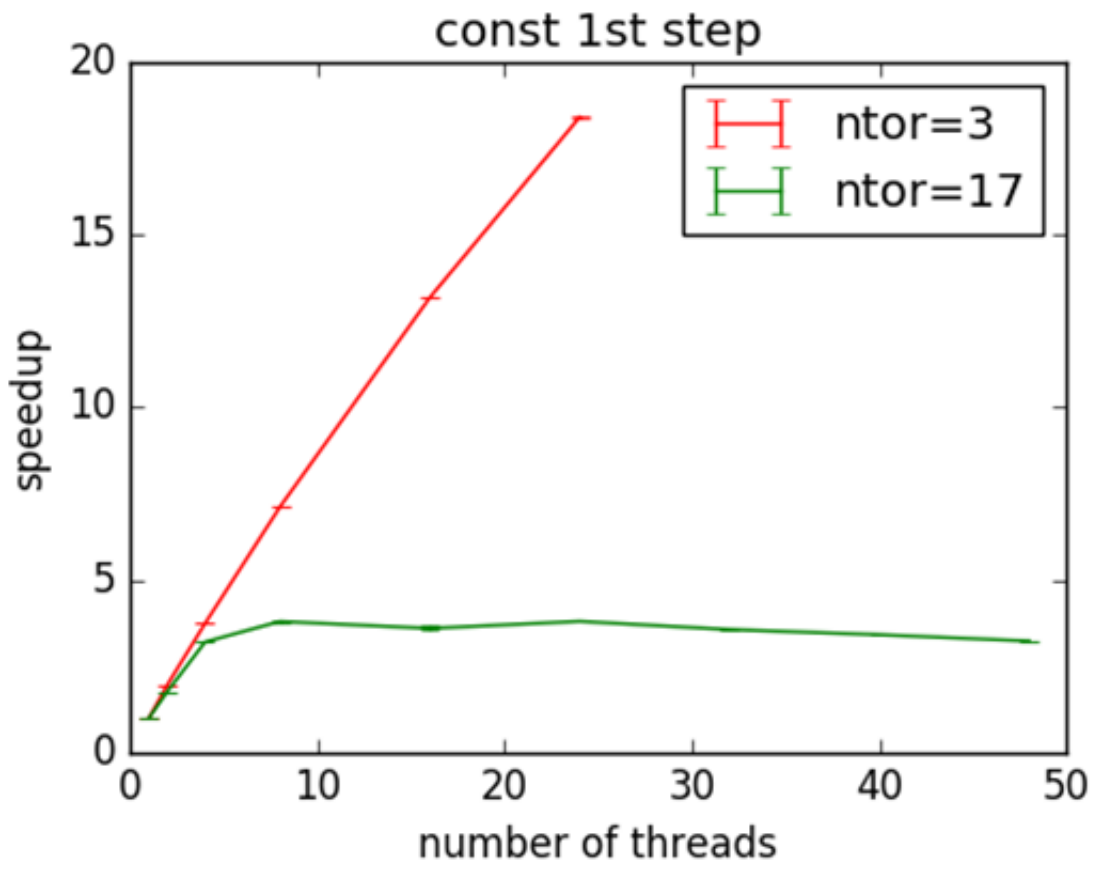

**Fig. 3** Speedup of the matrix construction for the small test case (red) and the medium size test case (green) on a Skylake node. The problem size is kept constant (strong scaling).

Two different methods were implemented in order to improve the scaling:

- Atomic: using a single atomic directive to update `A_glob`.
- Critical buffer: each thread stores values in a local buffer with 1M elements, when buffer full: small critical section to update `A_glob`.

We should note that synchronization is not absolutely necessary to set `irn_glob` and `jcn_glob` since all threads set the same value.

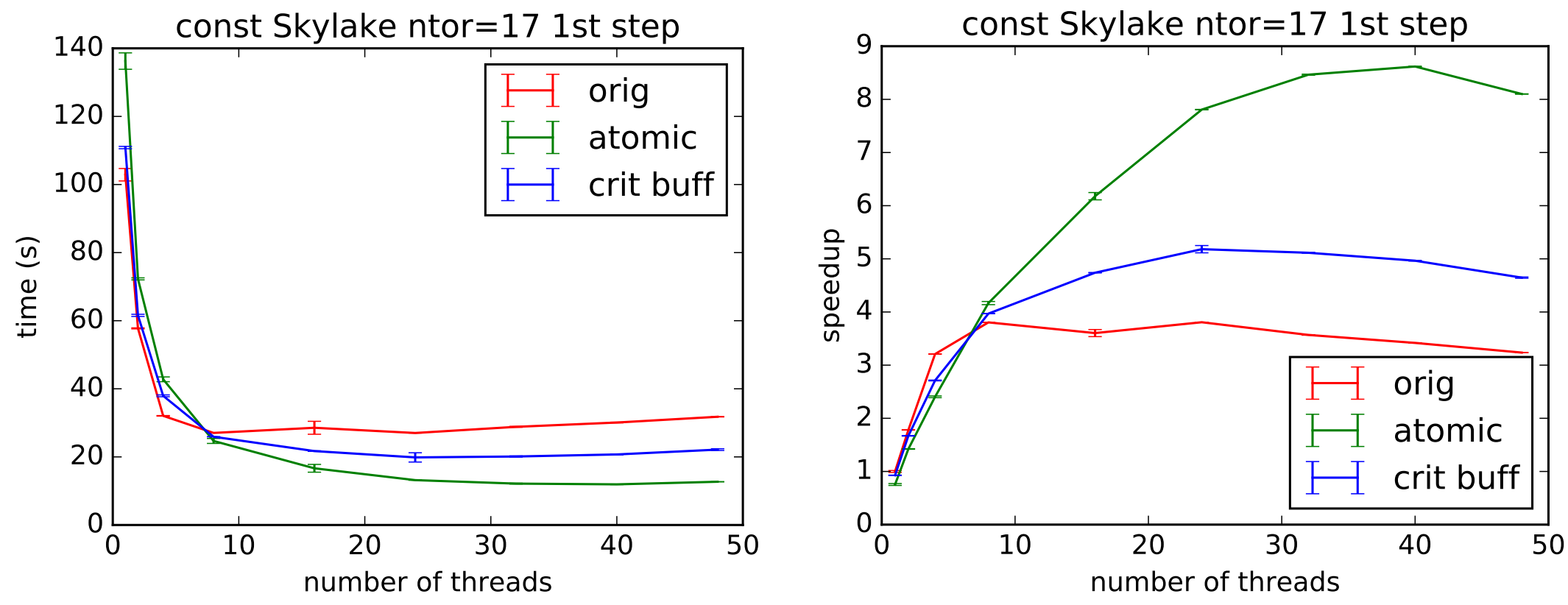

**Fig. 4** Matrix construction execution time (left) and speedup (right).

The right hand side of Fig. 4 shows that replacing the critical section with an atomic directive improves the speedup of the matrix construction by more than a factor of two. Introducing an extra buffer for the critical updates does not help significantly.

The left hand side of the figure shows that using the atomic directive leads to larger execution times on a single core. This is also visible on the right hand side of the figure (speedup), and it is later compensated by a better scaling properties. This will be further discussed in Section 5.6.

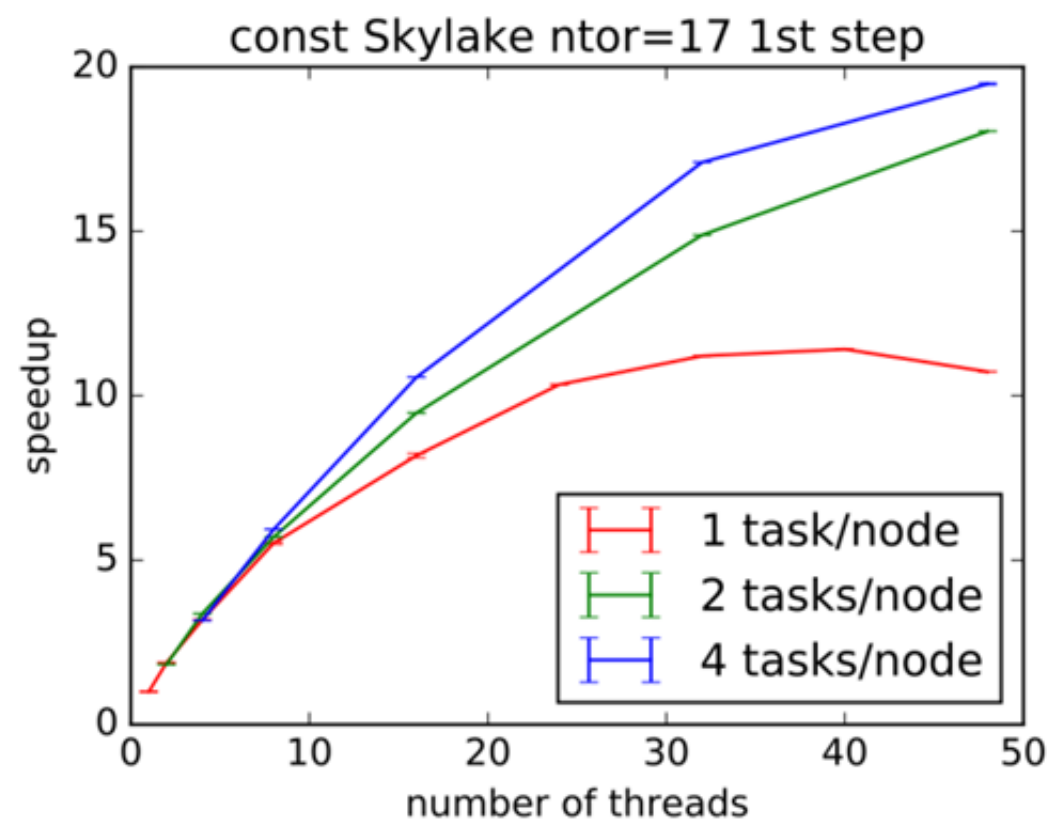

**Fig. 5** Scaling of the matrix construction using different numbers of MPI tasks. The execution time is shown as a function of the total number of threads/node.

The scaling of the matrix construction degrades above 24 threads. To improve this, we can divide the threads among more MPI tasks. This leads to less OpenMP synchronization and faster matrix construction as shown in Fig. 5.

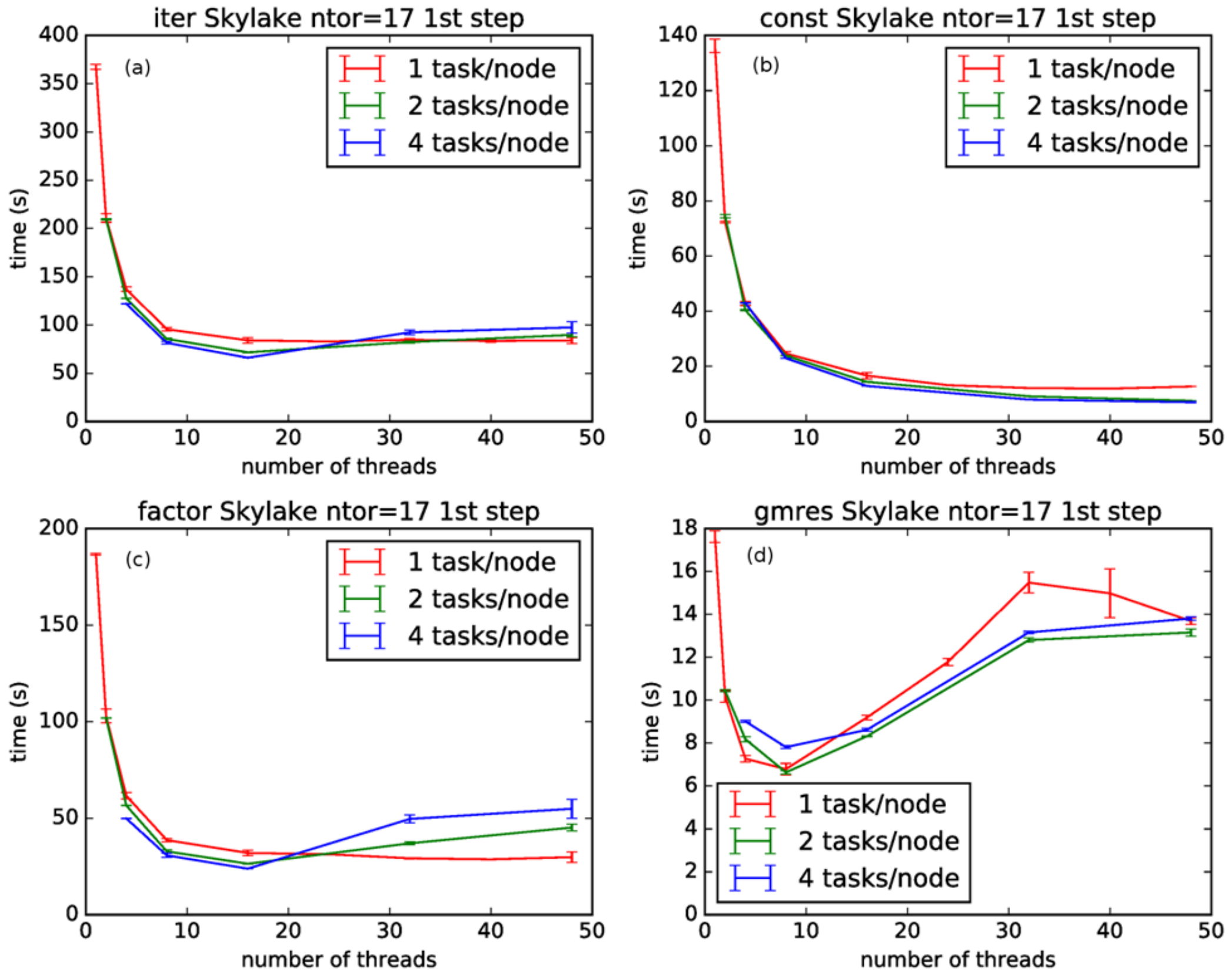

**Fig. 6** Execution times during the iteration after restart. The *x*-axis represents the total number of threads per node. (a) Total time of the iteration, (b) matrix construction, (c) LU factorization, (d) iterative solver.

The LU factorization and GMRES steps do not necessarily improve when we increase the number of MPI tasks (see Fig. 6). In fact, when all the cores of the nodes are utilized for the calculation, then the fastest setting for the PaStiX library is one task per node (see Fig. 6(c)). But if we limit the total number of cores per node that are used, then the factorization and solve steps are fast even with 4 MPI tasks (see Fig. 6(c)–(d) between 8 and 16 threads). A switch was implemented in JOREK which allows us to control the total number of threads used by PaStiX. We can define the following variable in the input file

```
pastix_maxthrd=16,
```

this would limit the total number of threads within a node that PaStiX can use to 16. The matrix construction keeps the value defined by *OMP_NUM_THREADS.* This leads to a factor of 1.7x speedup overall using 4 MPI tasks/node for the medium test case (72 MPI tasks in 18 nodes).

## 4.2. *KNL performance*

The modified code was tested on KNL nodes with different numbers of MPI tasks/node and different numbers of hyper-threads. The best configuration was found to be 4 MPI tasks per node, 17 threads/task for matrix construction, 4 threads/task for PaStiX and no hyper-threads. The execution times of the medium test case between KNL and Skylake are compared in Fig. 7. The code is 2.2–2.5 times faster on a Skylake node.

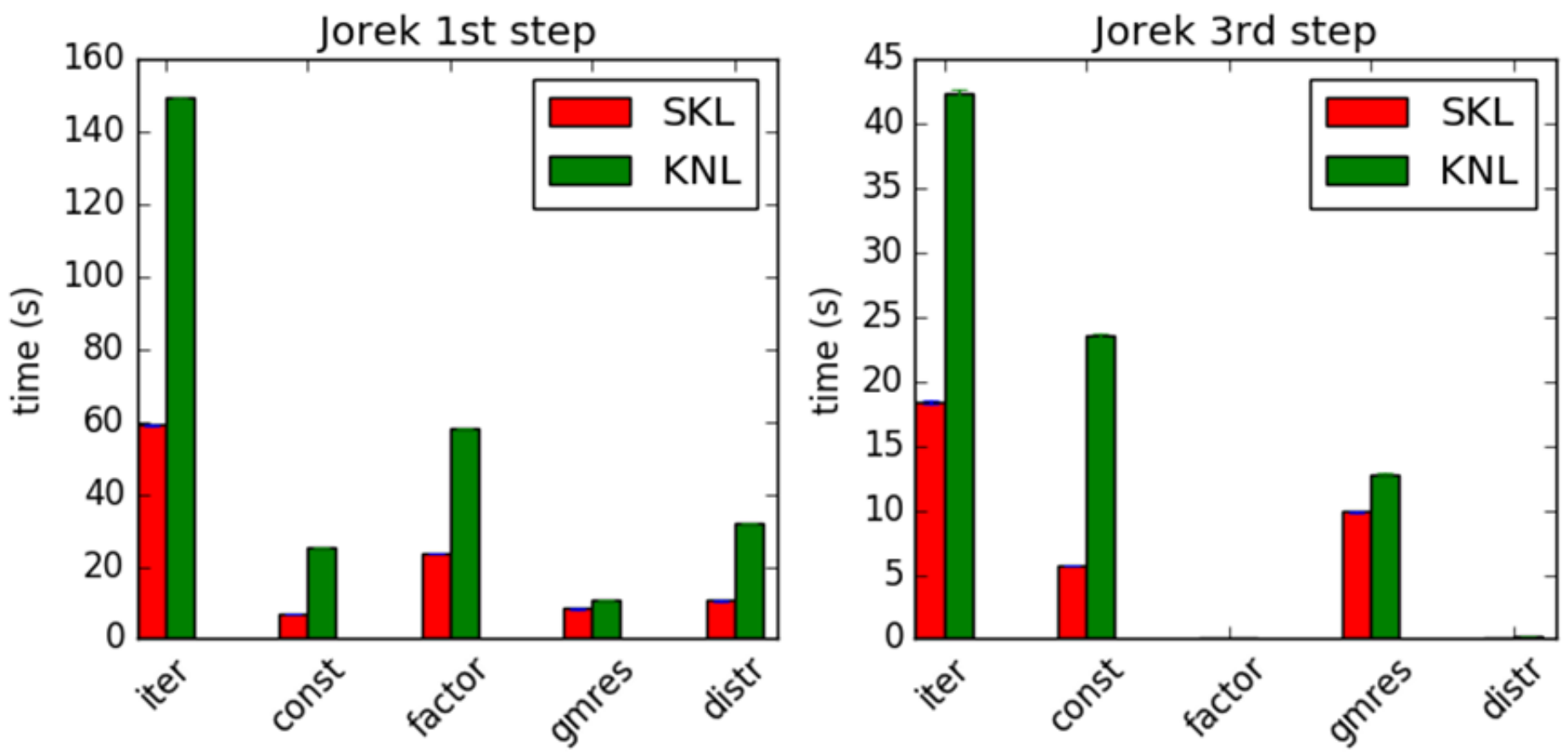

**Fig. 7** Medium test case (*ntor*=17) execution times on KNL and SKL

A few issues were observed with the medium size test case: PaStiX by default tries to use the *hwloc* library to bind the threads to the cores. This fails in some cases, causing very long execution times. Introducing an upper limit for the threads helps to overcome this problem.

# 5. **Vectorization**

The PaStiX library uses MKL routines, which should already be properly vectorized for the target architectures. In contrast, the matrix construction consists of user code only, and is not vectorized. To improve the performance of the matrix construction, the single core performance of the *element_matrix_fft* subroutine was investigated.

It requires 18 nodes to test the whole code with the medium size test case (*ntor*=17). In this section we are interested in single core and single node performance, therefore a testbed was prepared to call the *element_matrix_fft* subroutine independently of the main code. This allows us to perform single node tests of the subroutine with the same parameters (*ntor*=17). Unless otherwise mentioned, the tests in this section are performed on the Skylake partition using the Intel Fortran compiler.

## 5.1. *Matrix element calculation*

Fig. 8 shows an outline of the matrix construction subroutine. In lines 1–14 (compute loops) we have seven nested loops that define the *ELM_p* temporary array, by first calculating the intermediate values *A, B, C*; and then the elements of the *ELM_p* array. Each of *A, B,* and *C* refer to a larger set of scalar and array variables. These loops are computationally intensive since calculating *A, B,* and *C* involves around a thousand lines of arithmetic instructions. The iteration space of the individual loops is small, *n_gauss*, *n_vertex_max*, and *n_order+1* are all equal to four. The parameter *n_plane* is set in accordance with the physical model that we are calculating (Table 1). For the medium size test case *n_plane=32*.

In lines 15–21 in Fig. 8 (transform loops), the data elements from the *ELM_p* and three similar additional arrays are Fourier transformed and the result is stored in the *ELM* array. Considering the $O(n \log n)$ complexity of the fast Fourier transform with *n_plane=32* elements at a time, we can see that the arithmetic intensity of the transformation is only $\frac{32\log(32)}{32 \times 8} \approx 0.6 \frac{\text{Flops}}{\text{byte}}$, which makes it very likely that this part of the computation will be limited by the memory bandwidth.

compute loops

```
 1. do ms=1, n_gauss
 2.   do mt=1, n_gauss
 3.     do mp=1,n_plane
 4.       A(mp, ms, mt) = …
 5.       do i=1,n_vertex_max
 6.         do j=1,n_order+1
 7.           B(i,j,ms,mt) = …
 8.           do k=1,n_vertex_max
 9.             do l=1,n_order+1
10.               C(k,l,ms,mt) = …
11.               ij = idx(i,j)
12.               kl = idx(k,l)
13.               ELM_p(mp,ij,kl) = A(…)*B(…)*C(…)
14.   end do(s)
```

transform loops

```
15.   do i= 1,n_vertex_max*n_var*(n_order+1)
16.     do j= 1,n_vertex_max*n_var*(n_order+1)
17.       tmp=fft4(ELM_p(:,i,j))
18.       do k=1,n_tor/2
19.         do m=1,n_tor/2
20.           ELM(i*n_tor+k,j*n_tor+m) += tmp(k+m)
21.   end do(s)
```

**Fig. 8** Illustration of the matrix element calculation. Lines 1–14 calculate the *ELM_p* temporary array, lines 15–21 transform these and store the results in the *ELM* matrix.

## 5.2. *Specifying the memory layout of arrays*

We would like to rely on the auto vectorization of the compiler to improve the performance, but unfortunately the compiler is not able to vectorize the code in the original form. According to the optimization report (generated using the *-qopt-report=5* flag), the compiler does not have sufficient information about the following topics:

- aliasing of pointer arrays,
- array alignment,
- stride for multidimensional arrays,
- read/write dependency of temporary variables.

We will address the first three points in this section. In JOREK, a derived type is defined that wraps all the buffers which are needed during matrix element calculation:

```
type thread_buffer
  real, dimension(:,:), pointer :: ELM
  real, dimension(:,:,:), pointer :: ELM_p
  ! …
end type
```

These buffers are defined as pointers, which can decrease the performance because the compiler has to consider that the pointers could be aliased. Adding the *contiguous* attribute to the variables improves the execution time, but it is better to avoid pointers altogether. The only rationale for using pointers would be that pointers keep the array bounds when they are passed as assumed shape dummy arguments, but this feature is not used in JOREK. Therefore, we can safely change them to allocatable arrays.

An array of *thread_buffer* objects is allocated in the variable *thread_struct*, and thread *tid* has its buffers in *thread_struct(tid).* To avoid writing long expressions, these buffers are renamed in the *element_matrix_fft* subroutine by introducing new pointers:

```
real, dimension(:,:,:), pointer :: ELM_p
ELM_p => thread_struct(tid)%ELM_p
```

```
! do calculation using ELM_p.
```

This again poses difficulties for the optimizer. An elegant solution to avoid pointers for renaming would be the associate construct from Fortran 2003. Unfortunately, it is a rarely used feature and the Intel compiler cannot optimize it properly. A third option is to pass the buffers as subroutine arguments

```
call element_matrix_fft(…, thread_buff(tid)%ELM_p)
```

and to choose a short dummy argument name for the buffer:

```
subroutine element_matrix_fft(…, ELM_p)
#define DIM1 n_vertex_max*n_var*(n_order+1)
   real, dimension(n_plane, DIM1, DIM1) :: ELM_p
end subroutine
```

Dummy argument renaming was already applied to the *ELM* array in the original code, now it has been extended to 16 more arrays.

Passing the buffers as dummy arguments allows us to specify the array shape explicitly. This information is known at compile time and it helps the compiler to reason about the stride between different array elements, and about the alignment of the different columns of the arrays.

In order to specify the alignment of the start of the arrays, we use the *–align array64byte* compiler flag, which will ensure that all local and module variables are aligned at 64-byte boundaries. This way all our arrays are aligned correctly. We still have to add the

```
!$DIR ASSUME_ALIGNED ELM_p:64
```

compiler directive to indicate that the dummy arguments are also aligned. This information is not known to the compiler, because it depends on how exactly the function is called from other compilation units.

The previous steps inform the compiler about array alignment, array size, the stride of memory access, and ensure that no arrays are overlapping. These changes improve the performance by roughly 20% as shown in Fig. 9. We should note that most of the code is still not vectorized, for which the main reason is the assumed dependency between different variables. In the next section, we will investigate this point.

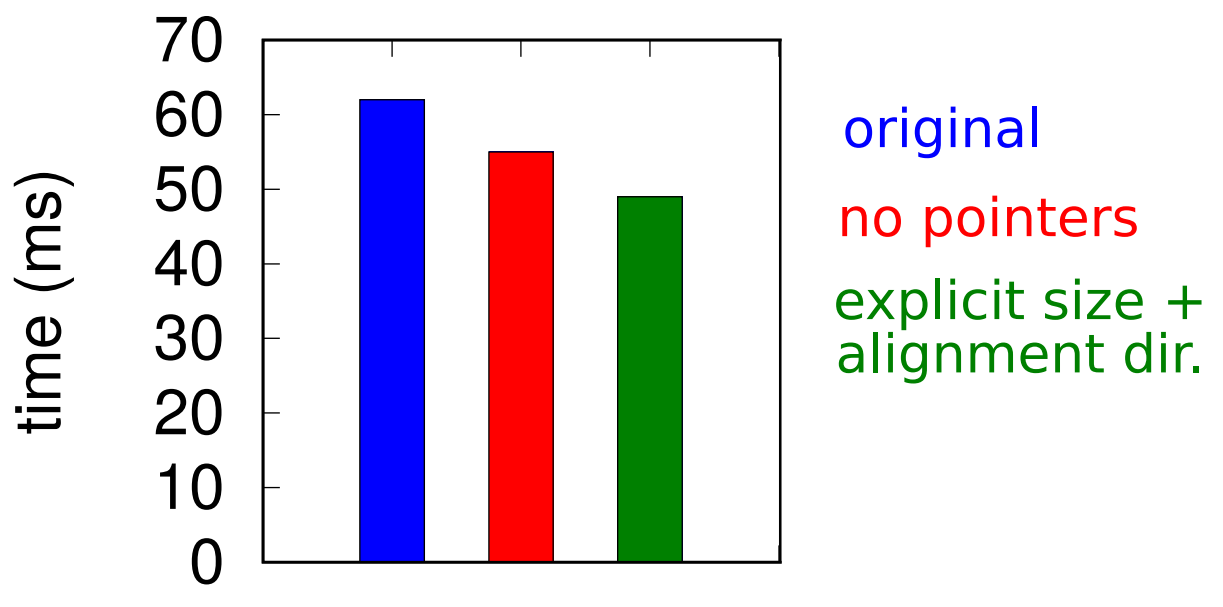

**Fig. 9** The effect of different optimization techniques on the runtime of the subroutine *element_matrix_fft*. The red bar shows the performance when pointers are avoided. The green bar represents the results when the explicit array size and the alignment directive is specified for the temporary buffers (and no pointers are used).

After ensuring that the compiler has all the information about the memory layout of the arrays, we can now investigate the performance of the compute loops and the transform loops individually. Fig. 10 shows that two thirds of the execution time is spent in the compute loops part. The next section will focus on these loops.

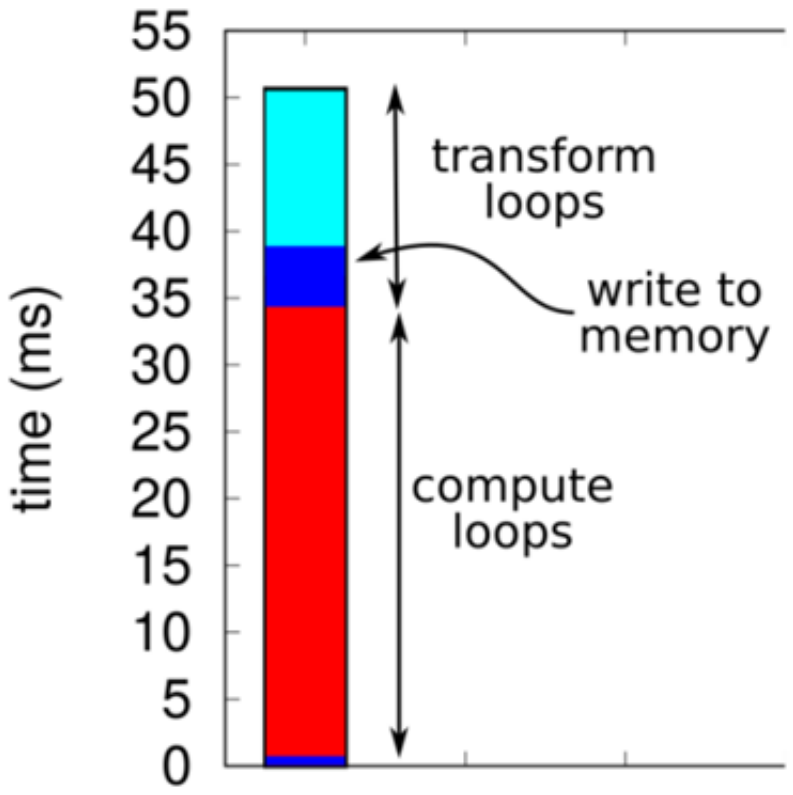

**Fig. 10** The execution time of the two main groups of loops in the matrix element calculations subroutine.

### 5.3. *Vectorization of the compute loops*

The changes listed in the previous section can already help the compiler to vectorize simple kernels. Unfortunately, the *element_matrix_fft* subroutine is too complex for the compiler to vectorize. The usage of hundreds of temporary variables creates several assumed dependencies, which the compiler cannot resolve.

The dependency problem is easy to fix with the *SIMD* directive from the OpenMP standard. Using the *private* clause, we can define the temporary variables to be private for each vector lane. The *SIMD* directive also tells the compiler that there is no vector dependency inside the loop and it can be safely vectorized.

The function calls inside the loops can also prohibit vectorization. To evaluate different vectorization strategies, the function calls were temporarily switched off.

Several ways to optimize the compute loops were tried. Fig. 11 shows the execution time of these different variants. The most promising loop for vectorization is the *mp* loop, which has an iteration space of 32. This is an outer loop, and we first tried to transpose it so that it becomes an inner loop. This slightly increases the execution time (see the bar labelled as *mp1*), because certain variables are calculated redundantly. After adding the *SIMD* directive the execution time decreases significantly. Using temporary arrays, we can avoid some of the redundant calculation, but it did not improve the execution time (*SIMD mp2*). It is also possible to vectorize the *mp* loop as on outer loop using the SIMD directive (*SIMD mp3*). Alternatively, the loops over the Gaussian points (*SIMD ms mt*) can be also vectorized if we use the collapse clause of the SIMD directive (otherwise the iteration count is too low). The last two variants have the best performance. Appendix 9 gives an overview of the different versions of the compute loops.

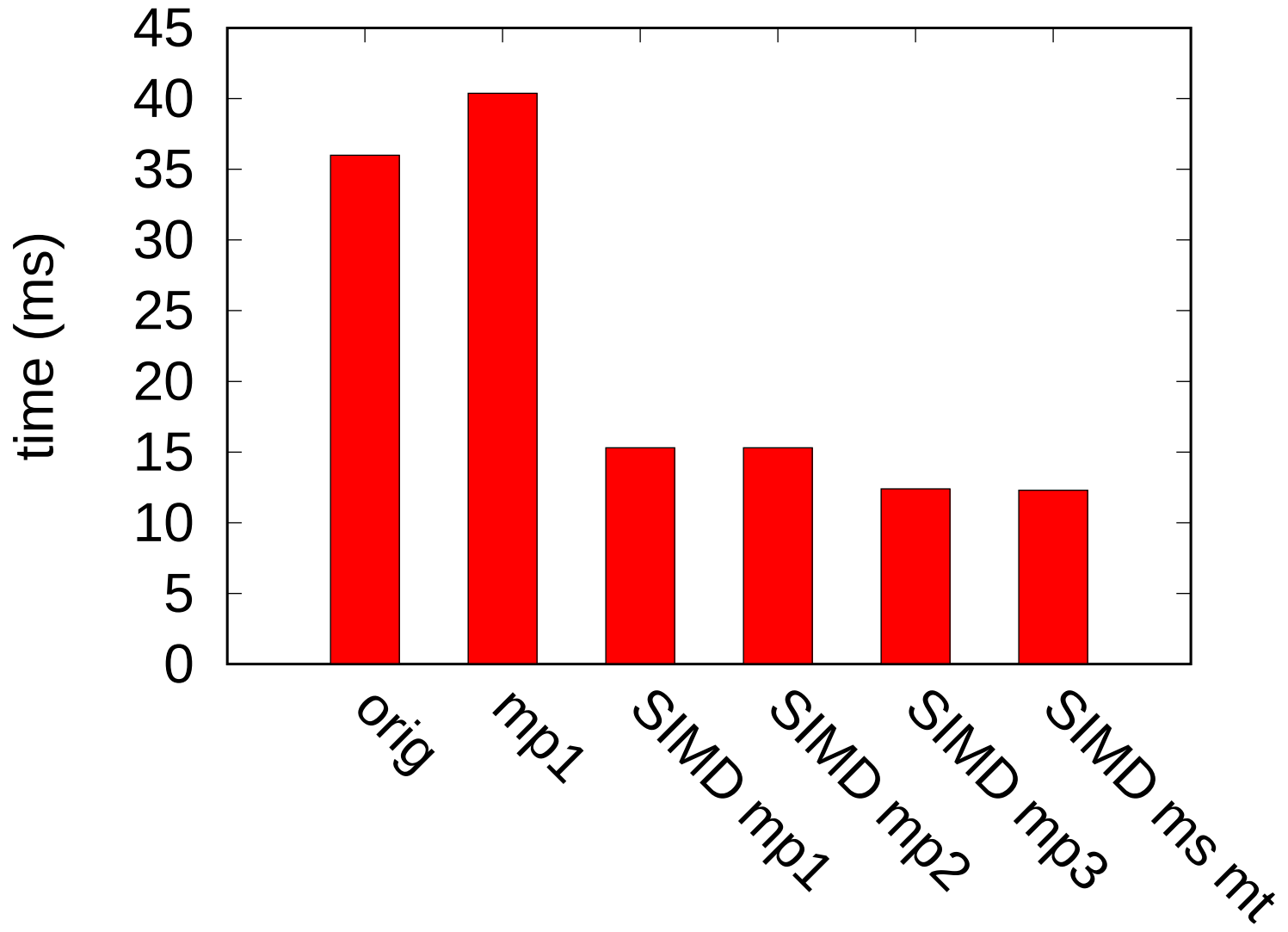

**Fig. 11** Execution time of the compute loops (without function calls) using different vectorization options: *orig* and *mp1* have no vectorization, *SIMD mp1-mp2* have a vectorized inner *mp* loop, *mp3* is the outer loop vectorization. The last bar is the vectorized version of the loops over the Gaussian integration points. In Appendix 9, the different versions of the loops are listed.

Using the outer *mp* loop vectorization, the compute loop nest is completely vectorized, and the *mp* loop is completely unrolled. Still we achieve only a factor of 3 speedup instead of the expected factor of 8. The explanation of this discrepancy is that we are processing a large amount of data (see next section) and after vectorization, the cache bandwidth becomes the bottleneck. According to Intel vectorization advisor, the performance is somewhere between the L2 and L3 cache bandwidth limit.

The compute loop nest contains a small number of function calls. For the measurements presented in Fig. 10, these function calls were switched off. These calls can be vectorized using the OpenMP *declare simd* directive. We have to provide the additional *–vecabi=cmdtarget* compiler flag, so that the compiler generates AVX-512 vector code. For the functions *corr_neg_temp* and *corr_neg_dens* inlining further helps the performance. We place these subroutines in a file that is included into the same compilation unit where the *element_matrix_fft* subroutine is defined. This way the compiler can inline these functions. An alternative option would be the *–ipo* compiler flag, but that decreases the overall performance.

So far, we have only tested single core performance. In Fig. 12 we present the execution time of the optimized subroutine as we increase the number of threads (weak scaling, work/thread is kept constant). Due to the optimization discussed so far, the subroutine is approximately 2.3x faster than the original. The optimized compute loop (blue bars) takes roughly half of the execution time of the matrix element calculation (blue + green bars). While the capacity to perform flops increases linearly with the number of cores, the memory bandwidth does not. The L3 cache is shared among the cores as well. That is why the subroutine performance deviates from the ideal scaling, which would be a flat constant line in this graph.

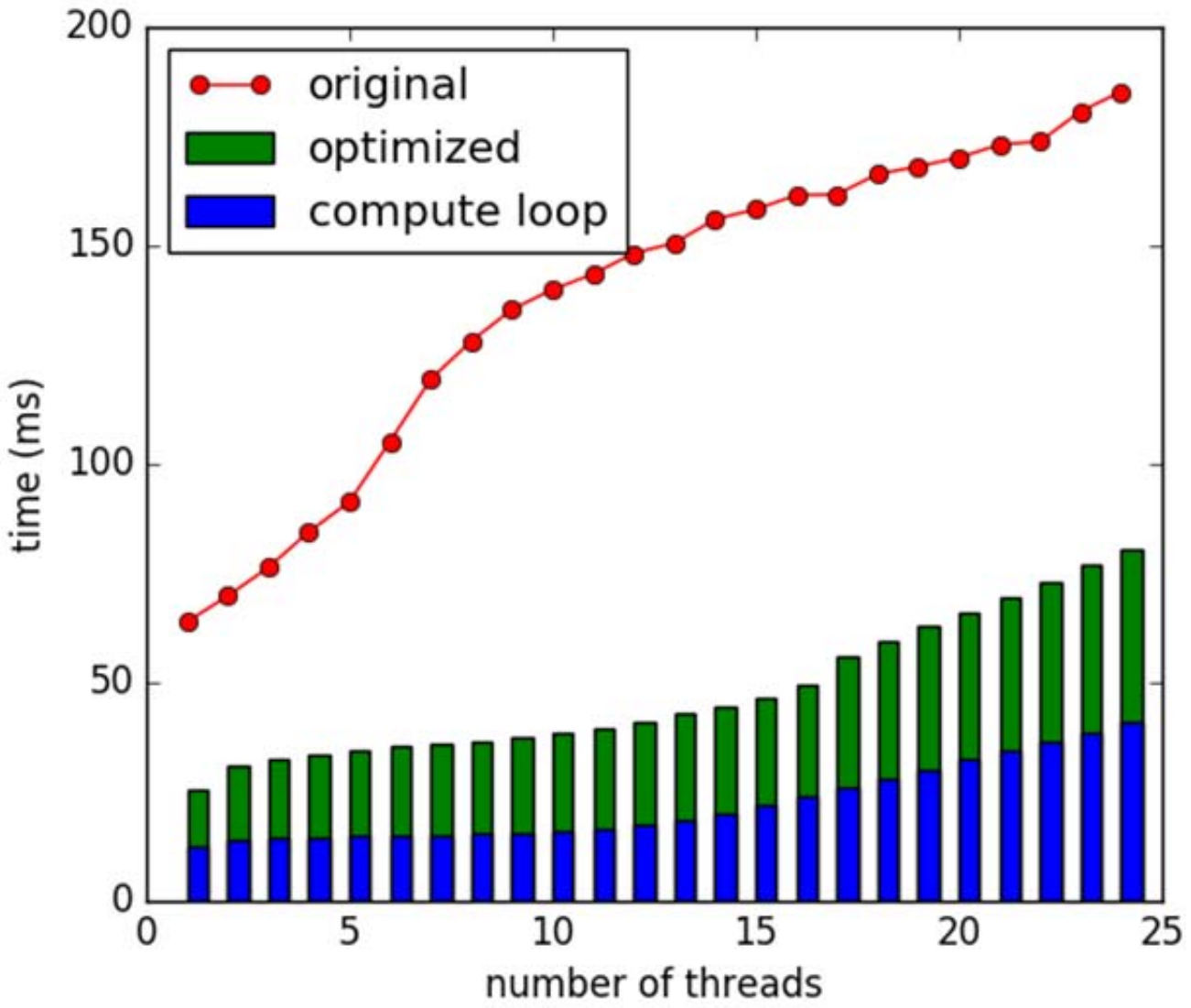

**Fig. 12** Weak scaling of the *element_matrix_fft* subroutine: the amount of work/thread is kept fixed, no synchronization between the threads, one thread/core is used. The red curve shows the execution time of the original code without any optimization. The bars show the execution time with the optimizations in the compute loops: the blue part is the execution time of the compute loop itself and the green bars shows the execution time of the rest of the subroutine.

## 5.4. *Memory and cache optimization*

The compute loop (lines 1–14 in Fig. 8) calculates *ELM_p* and three more temporary arrays: *ELM_n, ELM_k, ELM_kn*. The total size of these arrays is 4*3 MiB. The second loop nest (lines 15–21 in Fig. 8) takes the temporary results, applies a Fourier transformation, and stores the results in the *ELM* array. The four temporary arrays ELM_p, ELM_n, ELM_k, ELM_kn are processed using four loop nests to add the contributions from these arrays to the single *ELM* output array, whose size is 27 MiB.

We should note that the Skylake processor has 1 MiB L2 cache and 33 MiB L3 cache shared among the 24 cores. On KNL, 1 MiB of L2 cache is shared between two cores in a tile. In our kernel each thread processes 12 MiB of temporary data, and generates 27 MiB of output data, therefore the memory bandwidth and cache usage becomes an important factor.

The dark blue patch in Fig. 10 shows the measured execution time of writing the results to the memory. It takes around 4 ms, which is close to the theoretical estimate of writing the data (a modified stream benchmark shows 7 GiB/sec single thread write bandwidth in the Skylake partition).

In order to improve data locality while writing the results, the four individual loop nests were merged for the *ELM_p*, ELM_n,… arrays. The *i* and *j* loops were transposed and the inner loops that scatter the data from *tmp* to *ELM* were also transposed. These changes lead to a few ms speedup.

To improve cache usage, instead of filling the *ELM_p(:,:,:)* array completely (3 MiB) before adding it to *ELM(:,:),* we filled only the *ELM_p(:,:,i)* slice (192 kiB) and added it to *ELM(:,:).* This reduces the data size that needs to be cached by a factor of 16. The cache optimized subroutine has much better weak scaling properties, as we can see in Fig. 13. Using a single multicore CPU with 24 threads, the optimized code has a factor of four speedup compared to the original. During these optimization steps, the last two dimensions of the *ELM_p* arrays were transposed to have a favorable access pattern in the compute loops.

This is just an initial step in optimizing cache usage. The scaling of the compute loops (Fig. 12) indicates that there is still more room for cache and memory optimization.

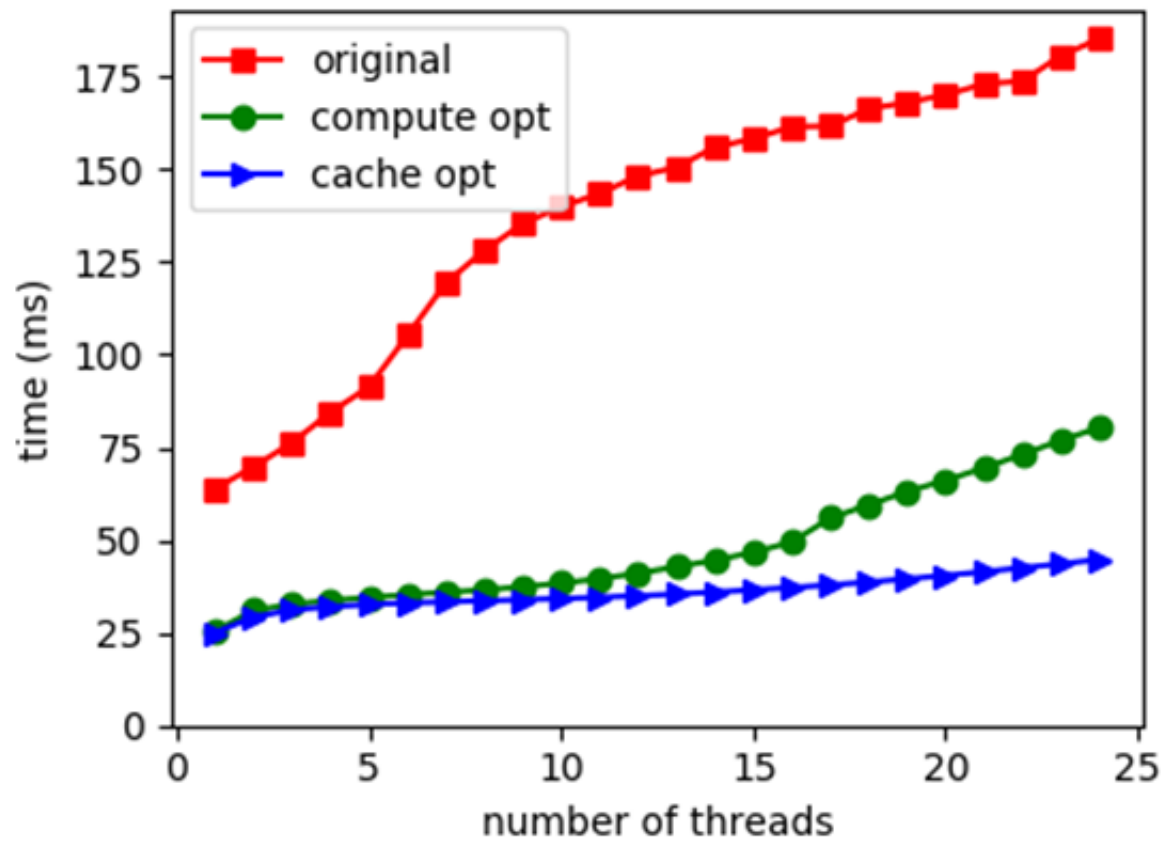

**Fig. 13** Weak scaling of the element matrix construction subroutine. The red and green curves show the same data as in Fig. 12. The blue curve adds cache optimization on top of it, which improves the scaling.

## 5.5. *Vectorization on KNL*

The KNL and Skylake architectures both have the AVX-512 vector instruction set. The optimizations presented in the previous section also improve the performance on KNL by a factor of three, as we can see in Fig. 14. The most important change in this case is the compute loop optimization. The cache architecture of KNL is different than the one of Skylake, and unfortunately KNL does not benefit much from the simple cache optimization that was described in the previous section.

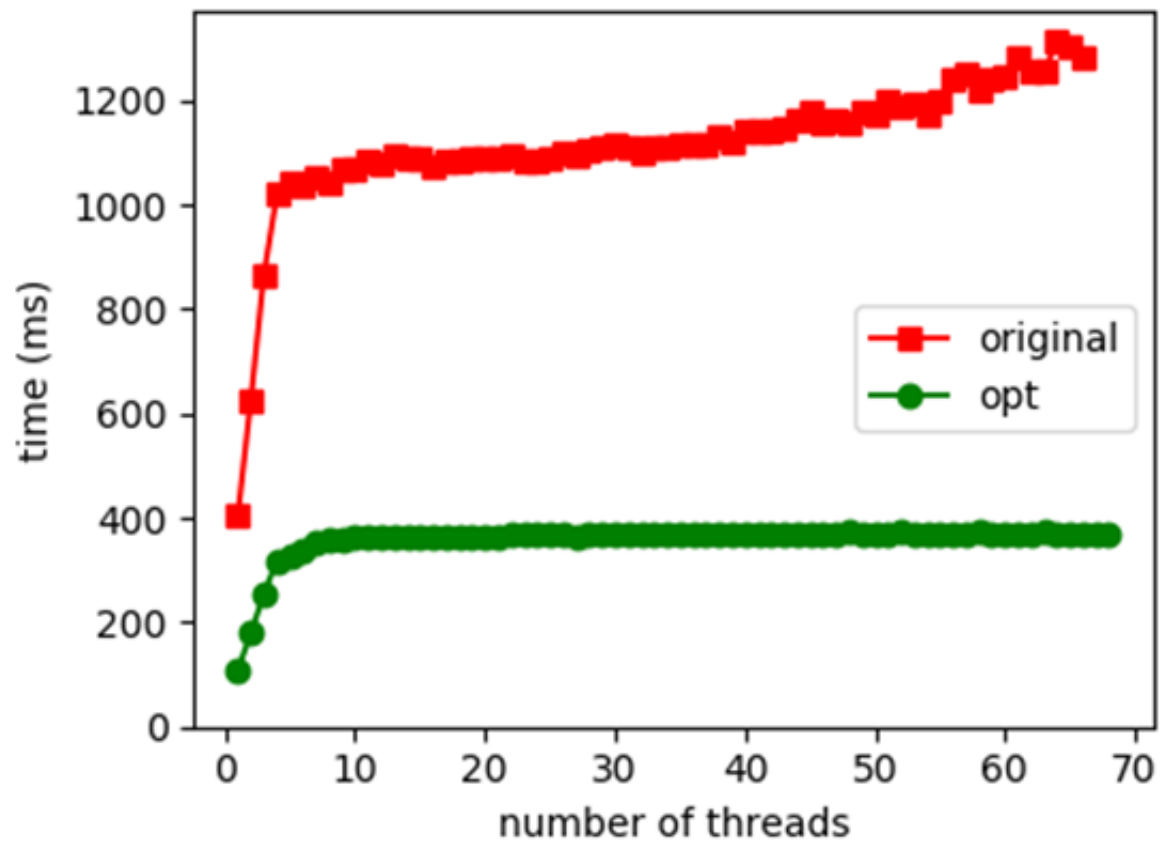

**Fig. 14** Weak scaling of the *element_matrix_fft* subroutine: the execution time is shown for the original (red) and the optimized (green) version.

## 5.6. *Integrating the changes into the main code*

The results presented in the previous sections were measured using a separate test program. After integrating the changes into the JOREK code, we measured the execution time of matrix construction in detail using VTune (Fig. 15). Comparing the purple bars between the left and right side, we can see that the execution time of *element_matrix_fft* is improved by a factor between 2.3x–3.6x (depending on the number of threads). The results from the *element_matrix_fft* subroutine are accumulated into the *A_glob* array. Updating *A_glob* needs synchronization among the threads. Using an atomic update (introduced in Section 4.1) has a large overhead even if we have only one thread (Fig. 15, left side). The execution time of the atomic synchronization is longer than the execution time of the optimized *element_matrix_fft* subroutine, therefore we considered additional implementations for the synchronization.

M. Hölzl proposed a variant of the synchronization using critical section and a temporary buffer, similar to the one presented in Section 4.1, but better aligned with the loop structure in the *construct_matrix* subroutine. The new synchronization method has no overhead for a small number of threads, and although it has a very long waiting time when we use a large number of threads (see orange bars in the right side of Fig. 15), it is still better then previous implementations. By choosing the number of MPI tasks and OpenMP threads appropriately, the new variant with critical section becomes faster than the atomic synchronization. The best execution configuration for the *ntor*=17 test case is to use 4 MPI tasks/node and 12 threads/MPI task on Skylake. This way the synchronization overhead is minimized, and the matrix construction is up to a factor of two faster (see Fig. 16).

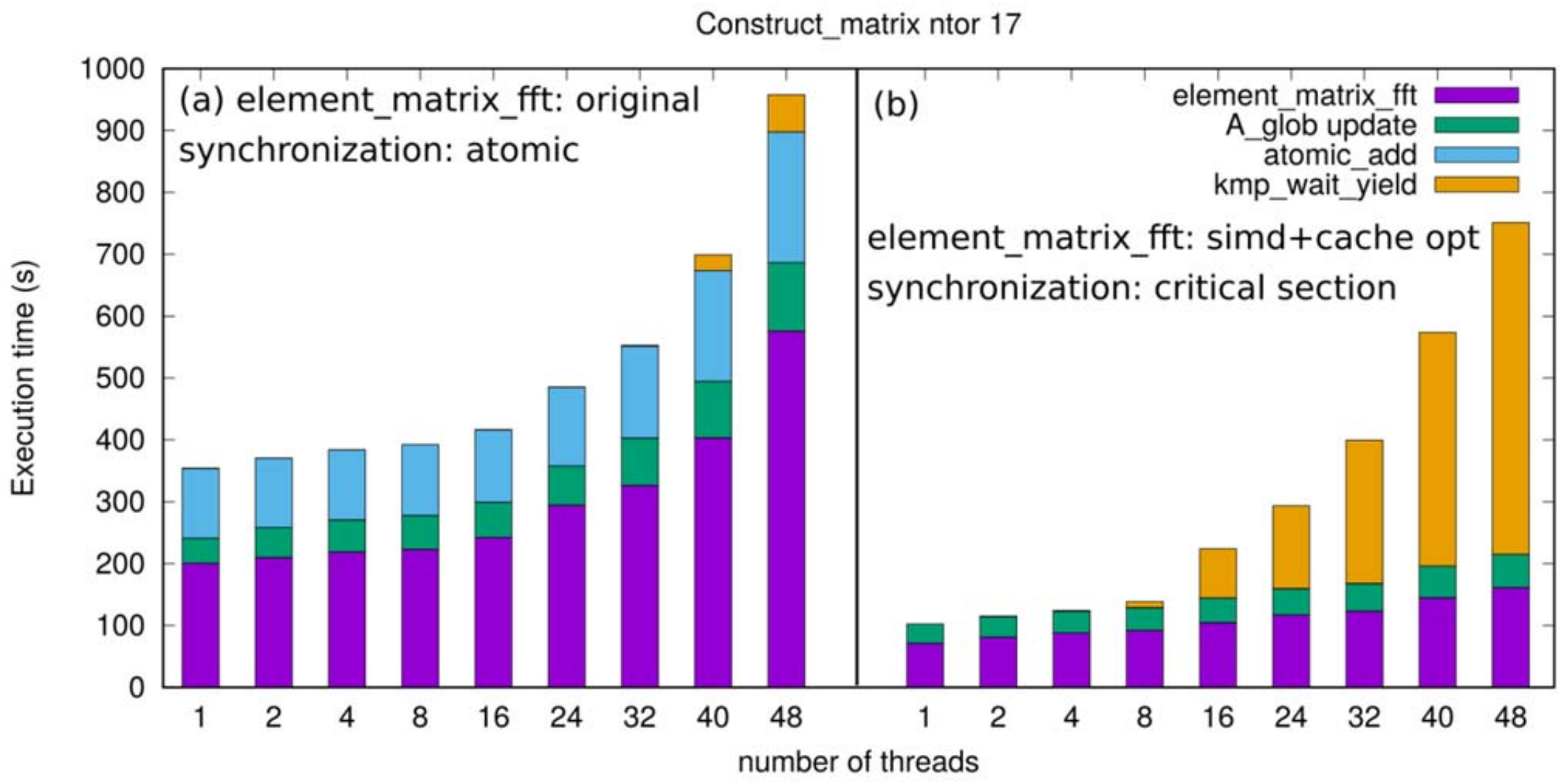

**Fig. 15** Execution time of matrix construction, the total execution time summed over all threads within a node shown (*ntor*=17 test case, 1 MPI task/node). Left side: original version of the *element_matrix_fft* and atomic synchronization, right side: optimized *element_matrix_fft* and improved critical section synchronization.

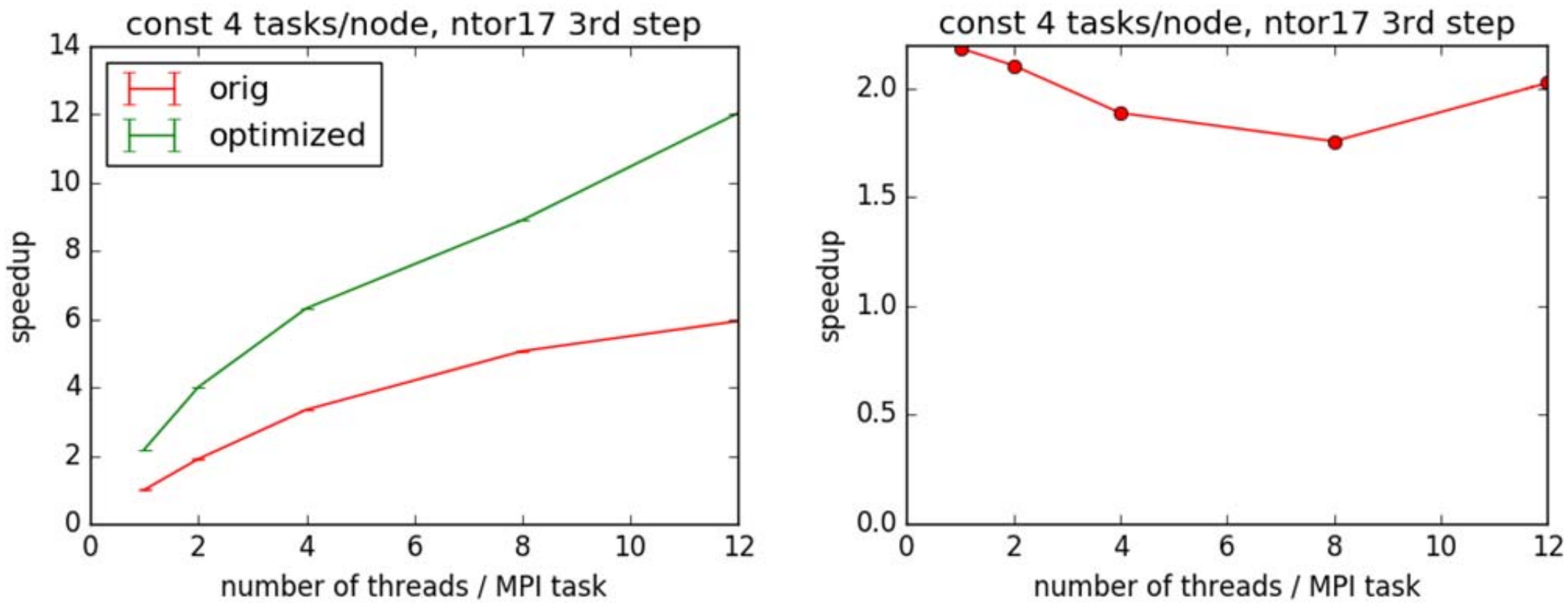

**Fig. 16** Strong scaling of the matrix (*ntor*=17 test case, 4 MPI tasks/node). The left side shows the speedup relative to the original code using one thread. The right side presents the ratio between the green and red curves.

We should note that there is one more OpenMP runtime function where a significant portion of the execution time is spent: kmp_wait_template. This subroutine has the same amount of execution time regardless of our changes of the synchronization method. Unfortunately, the profiling data collected by VTune was not adequate to reveal whether this function is called during matrix construction or not.

It might be possible to avoid synchronization during matrix construction altogether, since it is only the neighboring elements that write to the same memory locations. One could process the element in several batches, each batch set up in a way that their memory write pattern do not overlap. This idea is used in certain PIC codes during charge assignment step (Moschuering 2018).

# 6. **Summary**

The JOKLA project improved the performance of the JOREK code on the KNL and Skylake architectures. The OpenMP scaling of the matrix construction part of the code was investigated. Synchronization with atomic directive scales better for a large number of threads, but its overhead is comparable to the actual computation that needs to be synchronized. A modified version of the critical section synchronization together with a properly chosen hybrid MPI/OpenMP configuration gives the best performance.

It was found that the linear equation solver is also sensitive to the MPI/OpenMP execution configuration, and the best performance can be achieved if we limit the number of threads per node that is used by the solver.

The work continued by improving the vectorization of the matrix construction. In a separate test bed, the matrix construction subroutine was vectorized, and the data locality was improved, which lead to a 4x speedup for this subroutine on Skylake and a 3x speedup on KNL. After vectorization the code performance is limited by memory and cache bandwidth. The same changes integrated into the main code led up to a factor of two speedup of the matrix construction part. Future work should focus on further decreasing the synchronization overhead and the amount of copying the results between temporary buffers. The execution time of the JOREK code is generally 2.2–2.9 times longer on KNL than on Skylake.

# 7. **Acknowledgement**

This work has been carried out within the framework of the EUROfusion Consortium and has received funding from the Euratom research and training programme 2014-2018 under grant agreement No 633053. The views and opinions expressed herein do not necessarily reflect those of the European Commission.

# 9. **Appendix**

Different methods of vectorizing the *element_matrix_fft* subroutines are listed here.

| Name | Pseudo code | Remarks |
|---|---|---|
| MP1 | **do** ms=1, n_gauss; **do** mt=1, n_gauss<br>**do** i=1,n_vertex_max; **do** j=1,n_order+1<br>**do** k=1,n_vertex_max; **do** l=1,n_order+1<br>!$omp simd private(….)<br>**do** mp=1,n_plane<br>A(mp, ms, mt) = …<br>B(i,j,ms,mt) = …<br>rhs(mp,ij) = A(mp,ms,mt)*B(i,j,ms,mt)<br>C(k,l,ms,mt) = …<br>ELM(mp,ij,kl) = A*B*C<br>**end do(s)** ... | Loops transposed: *mp* loop in the innermost position.<br>Repeated calculation of *A* and *B*.<br>More than 300 SIMD private variables. |
| MP2 | **do** ms=1, n_gauss; **do** mt=1, n_gauss<br>**do** mp=1,n_plane<br>A(mp, ms, mt) = …<br>**end do**<br>**do** i=1,n_vertex_max; **do** j=1,n_order+1<br>B(i,j,ms,mt) = …<br>**do** mp=1,n_plane<br>rhs(mp,ij) = A(mp,ms,mt)*B(i,j,ms,mt)<br>**end do**<br>**do** k=1,n_vertex_max; **do** l=1,n_order+1<br>C(k,l,ms,mt) = …<br>**do** mp=1,n_plane<br>ELM(mp,ij,kl) = A*B*C<br>**end do**(s) | The *mp* loop is split into smaller parts.<br>Inner loop vectorization without repeated calculation.<br>Larger memory usage: 147 temporary scalars replaced with temporary arrays of dimension(32).<br>170 SIMD private variables. |
| MP3 | **do** ms=1, n_gauss; **do** mt=1, n_gauss<br>!$omp simd private(...)<br>**do** mp=1,n_plane<br>A(mp, ms, mt) = ...<br>**do** i=1,n_vertex_max; **do** j=1,n_order+1<br>B(i,j,ms,mt) = ...<br>rhs(mp,ij) = A(mp,ms,mt)*B(i,j,ms,mt)<br>**do** k=1,n_vertex_max; **do** l=1,n_order+1<br>C(k,l,ms,mt) = ...<br>ELM(mp,ij,kl) = A*B*C<br>**end do(s)** | Simple, original loop structure is kept.<br>More than 300 SIMD private variables.<br>Good performance. |
| MSMT | !$omp simd collapse(2) private(...)<br>**do** ms=1, n_gauss; **do** mt=1, n_gauss<br>**do** mp=1,n_plane<br>A(mp, ms, mt) = ...<br>**do** i=1,n_vertex_max; **do** j=1,n_order+1<br>B(i,j,ms,mt) = ...<br>rhs(mp,ij) = A(mp,ms,mt)*B(i,j,ms,mt)<br>**do** k=1,n_vertex_max; **do** l=1,n_order+1<br>C(k,l,ms,mt) = ...<br>ELM(mp,ij,kl) = A*B*C<br>**end do(s)** | Parallelization over the Gaussian integration points.<br>Loops collapsed to increase iteration count.<br>More than 300 SIMD private variables.<br>Non-unit stride memory access.<br>Good performance. |